\documentclass[review]{elsarticle}
\makeatletter
\def\ps@pprintTitle{To be published %
\let\@evenfoot\@oddfoot}
\usepackage{lineno,hyperref}

\usepackage{color}
\usepackage{ulem}
\usepackage{braket}
\usepackage{amsmath,amssymb,multirow,bm}
\usepackage{graphicx}
\usepackage{ulem}
\usepackage{cases}
\usepackage{comment}
\usepackage{graphicx,color}

\newcommand{\beq}{\begin{equation}}
\newcommand{\eeq}{\end{equation}}
\newcommand{\bea}{\begin{eqnarray}}
\newcommand{\eea}{\end{eqnarray}}

\begin{document}

\begin{frontmatter}

\title{Post-formation in alpha emission from nuclei}

\author[mymainaddress,mysecondaryaddress,mythirdaddress]{J. Tanaka\corref{mycorrespondingauthor}}
\cortext[mycorrespondingauthor]{Corresponding author.}
\ead{jtanaka@ikp.tu-darmstadt.de}
\address[mymainaddress]{Institut f\"ur Kernphysik, Technische Universit\"at Darmstadt, Darmstadt 64289, Germany}
\address[mysecondaryaddress]{GSI Helmholtz Center for Heavy Ion Research GmbH, Darmstadt 64291, Germany}
\author[3address,4address]{C.A. Bertulani}
\address[3address]{Department of Physics and Astronomy, Texas A \& M University-Commerce, Commerce, Texas 75429, USA}
\author[mymainaddress,mysecondaryaddress]{S. Typel}
\begin{abstract}
 We propose a novel mechanism to explain nuclear decay by emission of an alpha particle. We show that the famous Geiger-Nuttall law can be explained by {\it post-forming} an alpha particle outside the range of the nuclear interaction with the daughter nucleus. This contrasts with the commonly accepted mechanism of first alpha particle {\it pre-formation} followed by emission through barrier penetration. We predict that the post-formation mechanism is more likely to occur for $\alpha$-particles with higher energy.  
\end{abstract}


\end{frontmatter}

{\it Introduction}. The first application of the quantum mechanical concept of tunneling was made by George Gamow,  who in 1928 calculated the probability for an alpha-particle to tunnel through a Coulomb barrier and applied it to determine the lifetimes of nuclear alpha-decays \cite{gamow1928zur}. The calculations by Gamow were able to explain an empirical formula previously known as the Geiger-Nuttall law \cite{geiger1911lvii}:
\begin{equation}
{\rm{log}}_{10}t_{\alpha1/2} = aQ_{\alpha}^{-1/2} + b.
\label{eq.GN}
\end{equation}
Gamow's theory and its variations are still applied to determine not only alpha-decay lifetimes, but also other similar processes such as  fission. The decay constant of an unstable nucleus by alpha- (or any cluster-) decay follows the simple recipe described in the equation
\begin{equation}
\lambda_\alpha  = {\ln 2 \over t_{\alpha1/2}} = f_{pre}\times \nu P_{tun},\label{lambda}
\end{equation}
where $f_{pre}$ is the preformation factor of the alpha-particle at the nuclear surface, $\nu$  is the assault frequency with which it hits the internal edge of the Coulomb barrier, and $P_{tun}$ is the probability (penetrability) that it emerges outside the barrier. The preformation factor $f_{pre}$ is the least known factor in this theory. Theoretically, and to first order, it is proportional to the square of a overlap matrix element involving the wave function of daughter, emitted alpha-particle and parent nucleus. 

Intensive theoretical efforts have been carried out to quantify the magnitude of the  preformation factor $f_{pre}$.  The list of theoretical models is extensive; we refer to a recent review \cite{qi2018recent}. A theoretical microscopic description of the preformation factor is the most difficult aspect of the alpha-decay theory. The alpha particle has a diameter of about 3.4 fm, whereas medium to heavy nuclear alpha-emitters have a larger diameter, of about 10 fm. The difference is not so large and it is thus difficult to accept that alpha-particles can be treated as point-like particles, as often assumed in the literature in connection to Eq. \eqref{lambda}. The alpha preformation within a nucleus is certainly due to a subtle correlation involving the four nucleons inside the nucleus, and is also driven by the large alpha binding energy. Such many-body correlations may be very sensitive to the structure of the individual nucleus with a dynamical dissolution of the alpha-particle and a regrouping of nucleons. 

Currently accepted microscopic alpha-cluster theories suggest that the alpha clustering occurs at the nuclear surface. Because the nuclear density at the surface becomes small with the main contribution from weakly-bound nucleons, binding energy and be gained by forming alpha clusters locally at the surface \cite{typel2014neutron}.  It is also more difficult to keep the alpha particle intact in the dense and saturated core of the nuclei. A recent theoretical approach to alpha clustering in heavy nuclei using the generalized relativistic density functional model \cite{typel2010composition} studied the isotopic dependence of the number of alpha particles on the mass number of tin isotopes \cite{typel2014neutron}. It was found that the number of alpha particles decreases with increasing number of neutrons. The increase of the neutron skin thickness with a smaller probability of forming an alpha cluster due to the larger neutron-proton asymmetry at the surface. Hence it is expected that alpha clustering at the surface is directly correlated to neutron skins in nuclei. A recent experiment \cite{tanaka2019direct} confirmed this prediction by measuring the cross sections of the quasi-free proton-induced alpha knockout reaction on tin isotopes (from $^{112}$Sn to $^{124}$Sn). The isotopic dependence on the cross sections reflects the predicted isotopic dependence of the number of alpha particles in tin isotopes \cite{typel2014neutron,tanaka2019direct}. The extreme case of this phenomenon appears at the doubly magic self-conjugate tin nucleus $^{100}$Sn as a super-allowed $\alpha$ decay\cite{auranen2018superallowed}. 

\begin{figure}[ptb]
\begin{center}
\includegraphics[width=12cm]{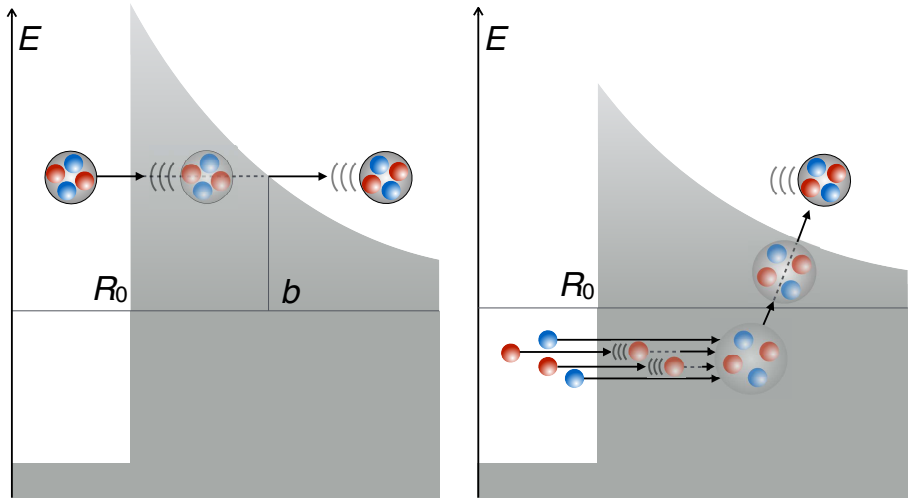}
\end{center}
\caption{Schematic view of the extreme situations in which an alpha-particle could escape a nucleus. {\it Left}: Tunneling of a single pre-formed alpha-particle  through the barrier. In this case, the alpha-particle is preformed through correlations arising from the interaction between nucleons inside the nucleus. {\it Right}: Tunneling of individual nucleons with the alpha-particle being post-formed outside the nucleus due to correlations of the four-nucleon interaction and the mean field of the daughter nucleus at the barrier region. The advantage here is that the Coulomb potential for protons is lower than that for $\alpha$-particles and there is no Coulomb barrier for neutrons.}
\label{fig1}
\end{figure}

{\it A novel approach to alpha-decay}. In this article we claim that the Geiger-Nuttall law can be explained with another mechanism, involving a ``post-formation" factor. It is well-known that tunneling of composite particles lead to subtle effects such as resonant tunneling, which appears when a matching occurs of energies in closed and open channels of a multiparticle system. It is a many-body generalization of the resonant tunneling of a particle through a barrier. 
This is used in practical applications such as in the resonant-tunneling diode \cite{tsu1973tunneling}. We show that instead of a pre-formation of the alpha-particle it is also possible that neutrons and protons individually tunnel and clusterize {\it outside} the Coulomb barrier. A schematic view of our hypothesis, as compared to tunneling of a preformed alpha, is shown in Figure \ref{fig1}. 

The new scheme proposed above leads to rich possibilities, such as deuterons being preformed inside the nucleus, tunnelling individually, and post-forming an alpha particle within or outside the Coulomb barrier. Other channels could include the  individual tunneling of $^3$He-n, or t-p pairs followed by alpha post-formation. Here we show that the celebrated  Geiger-Nuttall law also follows from such an alpha-post-formation theory. Our theory can be used to extract the probability of post-formation of the alpha-particles right outside the Coulomb barrier. Evidently, this post-formation probability is small as now several particles have to tunnel through the barrier. But nothing in nuclear structure theory disallows the possible existence of alpha correlations stemming from the tails of individual nucleon wave functions within or outside the Coulomb barrier.

{\it Extended Gamow model for alpha-particle constituents}. The theoretical problem of tunneling of composite particles is a formidable one. Even with  few particles, the theory has to deal with the rearrangement of  energy levels during tunneling, resonant tunneling, diabatic and adiabatic level crossings, the Landau-Zener effect, etc. \cite{bertulani2004electric,bertulani2007tunneling}. The probable reason why the Geiger-Nuttall law can be well described by the Gamow model for the tunneling of a point particle is the strong binding of the $\alpha$-particle. For loosely-bound systems, it has been proven that the compositeness of the particles have a strong influence on, e.g., nuclear fusion reactions \cite{canto2015recent}. Next we will prove that the Geiger-Nuttall rule also arises if one assumes tunneling of the $\alpha$-particle individual constituents followed by their clusterization at the nuclear surface. This raises intriguing questions associated with the meaning of alpha formation factors.

Our proof is relatively simple and therefore relates to universal characteristics of particle formation and decay in nuclei. Gamow used the semi-classical WKB approximation to calculate the barrier penetrability of an $\alpha$-particle with reduced mass $\mu_\alpha$ through a potential $V_\alpha(r)$.  The Coulomb potential for radii $r$ larger than the inner barrier radius $R$ is given by,
\begin{subnumcases}
{V(r)=}
V_\alpha(r)=\frac{2(Z-2)e^2}{r}\\
V_p(r)=\frac{(Z-1)e^2}{r}\\
V_n(r)=0,
\end{subnumcases}
for alpha particles, protons, and neutrons and a mother nucleus with charge number $Z$.
The penetrability $P_\alpha$ for an $\alpha$-particle  tunneling through this spherically symmetric Coulomb barrier is
\begin{equation}
\begin{split}
P_{\alpha}&\sim\exp\left\{-\frac{2}{\hbar}\int_R^b{dr{\sqrt{2\mu_\alpha(V_\alpha(r)-Q_\alpha)}}}\right\}\\
&\sim\exp\left\{-2\pi\frac{2(Z-2)e^2}{\hbar}\sqrt{\frac{\mu_\alpha}{2Q_{\alpha}}}\right\},
\end{split}
\label{eq.WKB}
\end{equation}
where $b$ is the outer turning point and the last line is valid for $b\gg R$.

We extend Gamow's model to deduce the total transmission coefficients  for the individual nucleons ($p,p,n,n$) composing the $\alpha$-particle. The penetration probabilities are given by the product of each probability, i.e.,
\begin{equation}
P_{ppnn}=P_{p}^2P_{n}^2.
\label{eq.PRODUCT}
\end{equation}
It is straightforward to show that the probability defined above has exactly the same form as the Geiger-Nuttall rule, Eq. \eqref{eq.GN}, if one uses the penetrability factors for individual nucleons as in the WKB  Eq. \eqref{eq.WKB} and the nucleon energies as a fraction, i.e., proportional to the Q-value, $Q_\alpha$. The same reasoning also applies by using and equation similar to  \eqref{eq.PRODUCT} for $\alpha$-particles post-formed by  individual tunneling of a deuteron, a proton, and a neutron (d + p +n), of two deuterons (d + d), of a triton and a proton (t + p), or of $^3$He and a neutron ($^3$He + n). The $Q_\alpha$ dependence of the Geiger-Nuttall rule is unaltered because the product of the individual probabilities  are factored out into a sum in the calculation of ${\rm{log}}_{10}t_{\alpha1/2}$. 

We can also use a slightly more rigorous method, starting with the radial Schr\"odinger equation for a nucleon $i$ with angular momentum $l$,
\begin{equation}
\frac{d^2\phi_{i}(r)}{dr^2}+\left\{\kappa^2-\frac{l(l+1)}{r^2}-\frac{2\mu_i}{\hbar^2}V_i(r)\right\}\phi_{i}(r)=0,
\label{eq.schroedinger}
\end{equation}
and the probability of barrier penetration to the nuclear radius $R$ is obtained from the regular $F_{l}$ and irregular $G_{l}$ solutions of the Eq.\eqref{eq.schroedinger} as,
 \begin{equation}
P_{l}=\kappa r\frac{\left|\phi^{+}_i(\infty)\right|^2}{\left|\phi^{+}_i(r)\right|^2}_{r=R}=\frac{\kappa r}{\left|F_l(\kappa r)\right|^2+\left|G_l(\kappa r)\right|^2}_{r=R},
\label{eq.BESSEL}
\end{equation}
where the $\phi^{+}_i(r)$ is the outgoing wave function for a nucleon and $\kappa$ is obtained from $E_{i}\sim(Q_{\alpha}-B_{\alpha})/4=\hbar^2\kappa^2/2\mu_{i} \ (<0)$, where $B_\alpha$ is the binding energy of $\alpha$ particle. This does not contradict the fact that for some nuclei $\alpha$ decay is observed and proton decay is not because of the energy conservation law.
This probability with $l=0$ yields for high-energy $\alpha$-particles the same result as Eq. \eqref{eq.WKB} in Gamow's semi-classical approach using positive $Q_{\alpha}$-values, $E_{\alpha}(=Q_{\alpha})=\hbar^2\kappa^2/2\mu_{\alpha} \ (>0)$.

Since the (bound) nucleons, or other alpha constituents, are assumed to be bound at large distances,   outside the range of the nuclear interaction with the daughter nucleus, we can use the asymptotic solutions of Eq. \eqref{eq.schroedinger},
\begin{equation}
\begin{split}
\phi_{p}(r)&=C_1\frac{W_{-\eta,l+1/2}(2\kappa r)}{r}\\
\phi_{n}(r)&=C_2\sqrt{\frac{2\kappa}{r}}K_{l+1/2}(\kappa r),
\label{eq.wavefunction}
\end{split}
\end{equation}
where the $W_{-\eta,l+1/2}$ is the Whittaker function, $\eta=\left(Z-1\right)e^2/\hbar v_p$, and $K_{l+1/2}$ is the modified spherical Bessel function. The coefficients $C_i$ are asymptotic normalization coefficients, which relate through normalization to the part of the single-particle wave function inside the nucleus.  The idea here is that the wave function tails of the bound nucleons post-form an $\alpha$-particle, which is emitted by acquiring a kinetic energy equal to the $\alpha$-binding energy minus the binding energy of the nucleons. Hence, nucleons close to the threshold would be favored.
Accordingly, the probability to find 2 protons and 2 neutrons at a large distance $r$ is  given by
\begin{equation}
P_{ppnn}(r)=C\left[\kappa rK^2_{l+1/2}(2\kappa r)\right]^2\left[W^2_{-\eta,l+1/2}(2\kappa r)\right]^2 . \label{ppnn}
\end{equation}
In fact,  the probability that the $\alpha$-particle is formed at the outer turning point,
$b=Z_\alpha Z_D\hbar c\alpha Q^{-1}_\alpha$ is  $f_{post}\times P_{ppnn}(b)$, where $f_{post}$ is the post-formation factor.
The half-life is inversely proportional to this probability, i.e., $t^{post}_{\alpha1/2} \sim \left[ f_{post} \nu_{p} P_{ppnn}(b)\right]^{-1}$. Instead of Eq. \eqref{ppnn}, we could also obtain $P_{ppnn}$ by solving the Schr\"odinger equation in the continuum to obtain the individual tunneling probabilities at the outer turning point, $r=b$. We have found that this procedure does not change the qualitative aspects of the results described below.  

\begin{figure}[h!]
\begin{center}
\includegraphics[width=12cm]{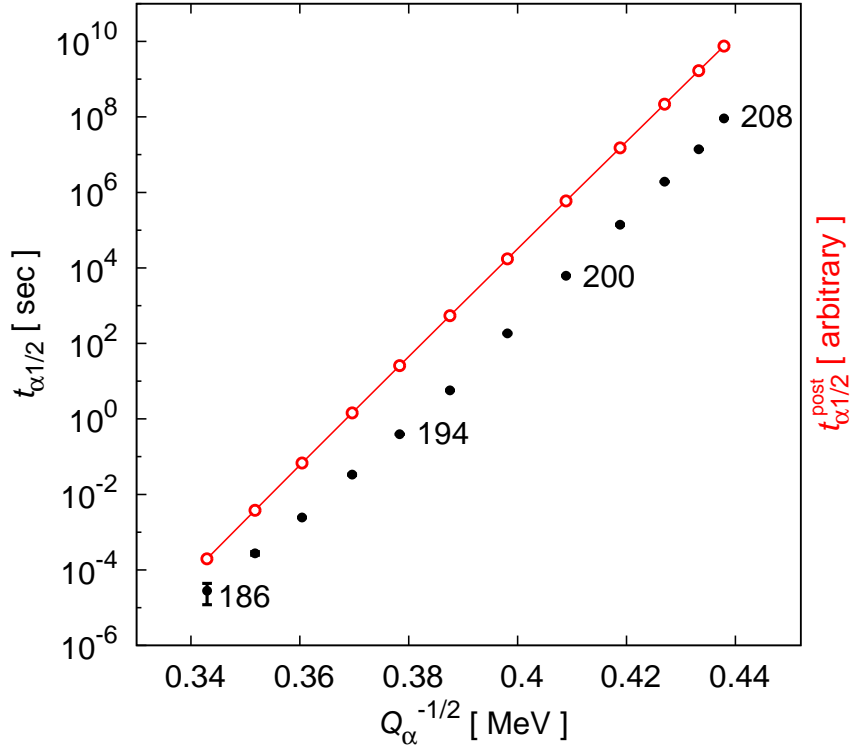}
\end{center}
\caption{Half-lives, $t_{1/2}$, of Polonium isotopes as a function of the inverse of the square root of $Q_{\alpha}$. The filled black points are the experimental data. The red open circles are the calculations with the post-formation hypothesis using formula \eqref{ppnn} for $ppnn$ individual tunneling case.}
\label{fig2}
\end{figure}

\begin{figure}[h!]
\begin{center}
\includegraphics[width=12cm]{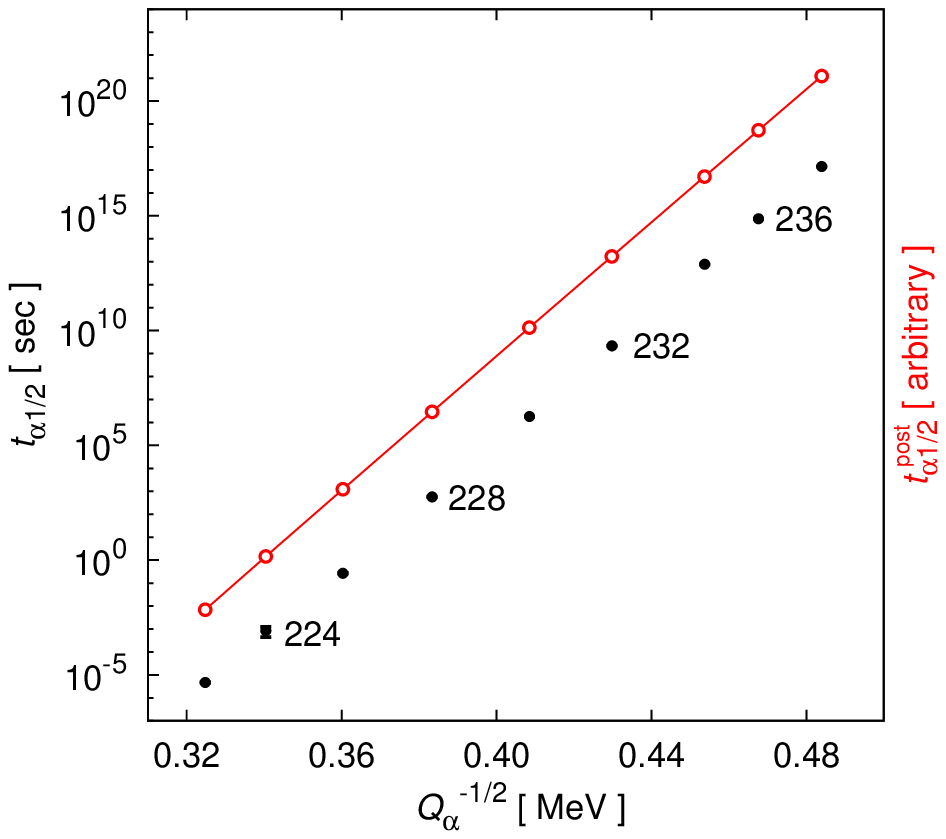}
\end{center}
\caption{Half-lives, $t_{1/2}$, of Uranium isotopes as a function of the inverse of the square root of $Q_{\alpha}$. The filled black points are the experimental data. The red open circles are the calculated half-lives from $P_{dd}$ with the post-formation hypothesis of $d$-$d$ pair tunneling. For the deuteron penetrability, the Whittaker function in Equation \eqref{eq.wavefunction} was used. $P_{dd}$ is the product of two deuterons' penetrabilities.}
\label{figU}
\end{figure}

{\it Numerical results}.
In Figure \ref{fig2} we show with red circles the results of a calculation based on the formula \eqref{ppnn} for the even Polonium isotopes (mass number $A=186 - 208$). For simplicity, we assume $l$=0 and $f_{post}=1$. The asymptotic normalization coefficients $C_i$ are obtained with a standard Woods-Saxon potential with $V_0$ obtained by solving the Shr\"odinger equation, range $R=1.2 A^{1/3}$ fm and diffuseness $a=0.65$ fm (see, e.g., \cite{huang2010radiative}). The potential depth is adjusted to reproduce the binding energy of the particle. Experimental values of $Q_{\alpha}$ values have been used. The black circles are the experimental data for the respective half-lives. One sees that there is indeed a remarkable similarity between the two results with regard to the dependence on $Q_{\alpha}$. In Figure \ref{figU}, we show the result of the calculation of half-lives of Uranium isotopes (mass number $A=222 - 238$) from $P_{dd}$ with the post-formation hypothesis of $d$-$d$ pair tunneling. Our results show that the tunneling of individual constituents of the $\alpha$-particle followed by a post-formation mechanism also reproduces the  Geiger-Nuttall law, i.e., the logarithm of the half-life is inversely proportional to the square root of $Q_{\alpha}$. Evidently, the assumption of $f_{post}=1$ is an overestimation and that is why our calculations yield larger half-lives than the experimental values. 
Physics intuition would lead to believe that $f_{post}\ll f_{pre}$.
Can we at least predict the order of magnitude of $f_{post}$ as compared to that of $f_{pre}$? Our answer is a resounding no, except if we could craft an accurate (ab-initio?) microscopic model for the post-formation factor. This seems to be a far-fetched hope with the theoretical techniques we have presently at hand. Only very rough estimates can be obtained such as using $t^{post}_{\alpha1/2}\sim t^{pre}_{\alpha1/2}$. In the case of post-formed $\alpha$-particles due to the fusion of two deuterons, this leads to 
\begin{eqnarray}
{f_{post}\over f_{pre}} &\sim& {\nu_{\alpha}P_{\alpha} \over \nu_{d}P_{dd}}  \label{fpostfpre}\\
&\sim& {\nu_{\alpha}\over \nu_{d}}\exp\left\{ - {4\pi Ze^2\sqrt{m_N}\over \hbar}\left[\sqrt{2\over Q_{\alpha}} - \sqrt{1\over Q_{d}}\right]\right\} ,\nonumber
\end{eqnarray}
obtained from Eq. \eqref{eq.WKB}, with $Z\gg (Z_{\alpha}, Z_{d})$, $\mu_{\alpha}\sim 4 m_{N}$, and $\mu_{d}\sim 2m_{N}$, where $m_{N}$ is the nucleon mass. We can use this equation to draw intuitive conclusions about the magnitude of the post-formation factor, with $Q_{d} \sim Q_{\alpha}-20$ MeV to account for the energy release as the deuterons fuse at the nuclear surface. For small values of $Q_{\alpha}$ ($\gtrsim  20$ MeV) the second square root within the exponential dominates, leading to $f_{post} \gtrsim f_{pre}$. 
For $Q_{\alpha}\sim 40$ MeV, $f_{post} \sim f_{pre}$, and for larger $Q_{\alpha}$ values $f_{pre}$ dominates over $f_{post}$. The known largest $Q_{\alpha}$ values are around 11 MeV for superheavy nuclei, but the $\alpha$ decay from the excited state of nuclei with large $Q_{\alpha}$ may follow this trend. 

{\it Summary and conclusions}. In summary, we have proposed a novel  mechanism to explain the Geiger-Nuttall law as the individual tunneling of nucleons, or other particles which can fuse to form an $\alpha$-particle at the nuclear surface. The results of our calculations indicate that the post-formation mechanism cannot be ruled out and might be significant to explain $\alpha$-decay.  The post-formation factor, $f_{post}$, is predicted to be largest at small values of $Q_{\alpha}$.  We suggest further that experiments exploring $\alpha$-knockout from radioactive projectiles studied in inverse kinematics along an isotopic chain are probably the best way to test if $\alpha$-particles are pre-formed in the nuclear interior or post-formed at the nuclear 
surface. More experimental information is desired to assess this long standing problem in nuclear physics: Where and how are $\alpha$-particles formed within a nucleus? We hope that this work stimulates further studies to answer this apparently simple question.

This work was supported in part by the U.S. DOE grants DE-FG02-08ER41533 and  the U.S. NSF Grant No. 1415656.   

\end{document}